# Intrinsic phase separation in superconducting K$_{0.8}$Fe$_{1.6}$Se$_2$ ($T_c$= 31.8 K) single crystals

Alessandro Ricci[1], Nicola Poccia[1], Boby Joseph[1], Gianmichele Arrighetti[2], Luisa Barba[2], Jasper Plaisier[3], Gaetano Campi[4], Yoshikazu Mizuguchi[5], Hiroyuki Takeya[5], Yoshihiko Takano[5], Naurang Lal Saini[1], Antonio Bianconi[1]

[1]Department of Physics, Sapienza University of Rome, P.le Aldo Moro 2, 00185, Rome, Italy

[2]Institute of Crystallography, National Council of Research, Elettra, 34012, Trieste, Italy.

[3]Elettra Synchrotron Center, MCX, 34012, Trieste, Italy.

[4]Institute of Crystallography, CNR, via Salaria Km 29.300, Monterotondo Roma, I-00015, Italy

[5]National Institute for Materials Science, 1-2-1 Sengen, Tsukuba, 305-0047, Japan

**ABSTRACT**

Temperature dependent single-crystal x-ray diffraction (XRD) in transmission mode probing the bulk of the newly discovered K$_{0.8}$Fe$_{1.6}$Se$_2$ superconductor ($T_c$ = 31.8 K) using synchrotron radiation is reported. A clear evidence of intrinsic phase separation at 520 K between two competing phases, (i) a first majority magnetic phase with a ThCr$_2$Si$_2$-type tetragonal lattice modulated by the iron $\sqrt{5} \times \sqrt{5}$ vacancy ordering and (ii) a minority non-magnetic phase having an in-plane compressed lattice volume and a $\sqrt{2} \times \sqrt{2}$ weak superstructure, is reported. The XRD peaks due to the Fe vacancy $\sqrt{5} \times \sqrt{5}$ ordering in the majority phase disappear by increasing the temperature at 580 K, well above phase separation temperature confirming the order-disorder phase transition. The intrinsic phase separation at 520K between a competing first magnetic phase and a second non-magnetic phase in the normal phase both having lattice superstructures (that imply different Fermi surface topology reconstruction and charge density) is assigned to a lattice-electronic instability of the K$_{0.8}$Fe$_{1.6}$Se$_2$ system typical of a system tuned at a Lifshitz critical point of an electronic topological transition that gives a multi-gaps superconductor tuned a shape resonance.

*Introduction* – The discovery of iron based superconductors (FeSC) made of FeAs superconducting layers intercalated by different spacer layers [1,2] has provided a new class of heterostructures at atomic limit where the lattice structure and its instability are key parameters controlling the complex Fermi surface topology giving multi-gap superconductivity stable at high temperature. Recently a new system A$_x$Fe$_{2-y}$Se$_2$ made of FeSe superconducting layers intercalated by spacer layers A= K [3,4], Cs [5,6], Rb [7], (Tl,Rb) [8], (Tl,K) [9] etc, has provided an additional system with a different Fermi surface topology. The most striking common feature of the FeSC is the presence of an active layer of FePn or FeCh (Pn: pnictogen and Ch: chalogen) edge sharing tetrahedrons. The A$_x$Fe$_{2-y}$Se$_2$ show multi-gap superconductivity with concentric multiple Fermi surfaces as  has been proposed in the shape resonance scenario for multi-gap superconductors made of heterostructures at atomic limit [10]. Heterostructures of metallic layers at atomic limit provide metals with mini-bands where high T$_c$ is controlled by the fine tuning of the chemical potential in the range of tens-hundreds meV around Lifshitz electronic topological transitions [10]. At such as low energy scale the fine tuning of the lattice structure of the superconducting active layer in these heterostructures at atomic limit is of high importance. The structure of the active superconducting layers is controlled by changing the spacer layer via the misfit strain between the active layer and the spacer layers as in cuprates [11]. The average crystal structure of the A$_x$Fe$_{2-y}$Se$_2$ at room temperature has already been established by different structural studies [12-19]. These studies revealed the structure of A$_x$Fe$_{2-y}$Se$_2$ to be of "122" type, with the presence of an Fe-square lattice decorated by regular array of vacancies [12-19] and hence a unit-cell structure five times larger than the basic ThCr$_2$Si$_2$-type tetragonal unit-cell.

In Fig. 1(b) shows a schematic view of the basic "122" type unit-cell and an expanded cell which can accommodate ordered vacancies at one of the Fe-sites. Temperature dependent neutron diffraction studies have revealed that the structural and magnetic phase transitions are related with the vacancy ordering in this system [13,14,18].

Here we present a temperature dependent single-crystal x-ray diffraction study of the $K_{0.8}Fe_{1.6}Se_2$ superconductor ($T_c$ = 31.8 K) to understand the structural dynamics of the system. We use high energy x-ray synchrotron radiation diffraction in transmission mode which allows us to probe the intrinsic bulk structure. We confirm previous works showing that the system undergoes an order-disorder transition at 580 K, as evidenced by the disappearance of the superstructure peaks due to the vacancy ordering. The superstructure peak intensities follow the same behavior upon heating and cooling with no detectable temperature hysteresis.

The main discovery of this work is that, unlike the earlier diffraction studies using powder samples [13,14], our high-resolution single-crystal x-ray diffraction data show the occurrence of a phase separation in $K_{0.8}Fe_{1.6}Se_2$ below 520 K (about 60 K lower than the vacancy ordering temperature) between a first in-plane expanded majority phase and a second in-plane compressed minority phase. Moreover, on decreasing the temperature below 520K a new set of superstructure diffraction peaks associated with the appearance of the minority phase is observed. These results constitute clear evidence of an intrinsic phase separation in 122 chalcogenides $K_{0.8}Fe_{1.6}Se_2$, which is assigned to the lattice instability of an electronic system near a Lifshitz critical point where a majority magnetic phase competes with the superconducting multi-gap phase where $T_c$ is amplified by a shape resonance [10]. The competition between the two phases provides a complex phase separation where both lattice structures show different lattice reconstructions.

*Experimental* - Single crystals of nominal composition $K_{0.8}Fe_{1.6}Se_2$ were prepared following the method described in Ref. [4]. The actual composition of the crystal was estimated to be K:Fe:Se=0.6:1.5:2 using an average of four points of the EDX measurements. The resistivity and magnetization studies showed the presence of a sharp superconducting transition at about 31.8 K ($T_c$ onset 33 K) [4]. The x-ray diffraction (XRD) data on the single crystal samples were obtained at the XRD1 beamline of ELETTRA synchrotron radiation facility in Trieste. The samples were oriented by means of a K diffractometer with a motorized goniometric X-Y stage head and a Mar-Research 165 mm CCD camera. The data were collected in transmission mode, with a photon energy of ~20 keV ($\lambda$ = 0.61992 Å), selected from the source by a double-crystal Si(111) monochromator. The x-ray diffracted beams were detected by a 2D CCD detector (MAR-Research), kept at a distance around 70 mm from the sample. Data from a $LaB_6$ standard was also collected for calibration. Measurements were conducted between 80 and 600 K with a temperature step of 3 K for the heating (80-600 K) and 2 K for the cooling (600-80 K) runs. For the measurements in the range of 80-300 K, sample temperature was varied and controlled by means of a cryo-cooler (700 series Oxford Cryosystems). For the measurements in the temperature range 300-600 K, a heat blower facility (Oxford Danfysik gas blower, DGB-0002) was used. In both cases, the temperature control was better than ±1 K. The single crystal x-ray diffraction images measured were properly processed using FIT2D program. The processed images were analyzed using a matlab® based software-package developed in-house.

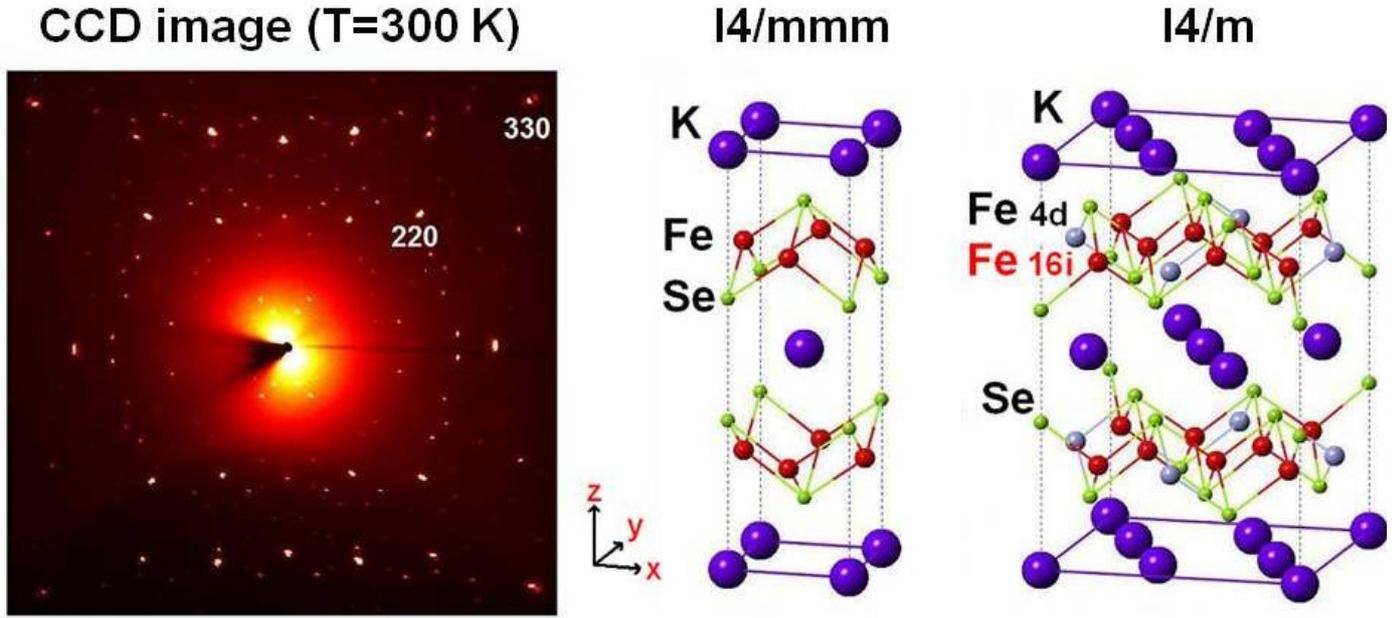

**Figure 1:** Single crystal x-ray diffraction pattern (CCD image) of $K_{0.8}Fe_{1.6}Se_2$ at 300 K. Structural models of the basic (I4/mmm) and larger (I4/m) unit-cell of the 122-type tetragonal structure are also shown.

*Results* – A typical single crystal XRD as-obtained pattern at 300 K is shown in figure 1. The pattern at room temperature is similar to the one reported in ref. [15] for a similar compound. The crystal displays a tetragonal $ThCr_2Si_2$-type diffraction pattern with overlapping superstructure peaks. Structural models corresponding to the basic and extended unit cells are also shown in Fig. 1. The extended unit-cell has two sites for the Fe (4d and 16i), which permits the vacancies to preferentially occupy one of the sites and order. The vacancy ordering observed at room temperature gives rise to the superstructure diffracted spots that can be indexed with a $\sqrt{5} \times \sqrt{5} \times 1$ expanded unit-cell (symmetry *I4/m*) of the basic $ThCr_2Si_2$ structure [13-19]. The a, b, lattice parameters, corresponding to the basic unit-cell are determined from the pattern shown in figure 1. The obtained value is $a = b \approx 4.01(3)$ Å. A careful observation of the image shown in figure 1 reveals that the principal diffraction spots are surrounded by superstructure peaks forming a group of eight reflections (see inset in Fig. 2 lower panel) in agreement with the data reported for similar systems at room temperature [15,18].

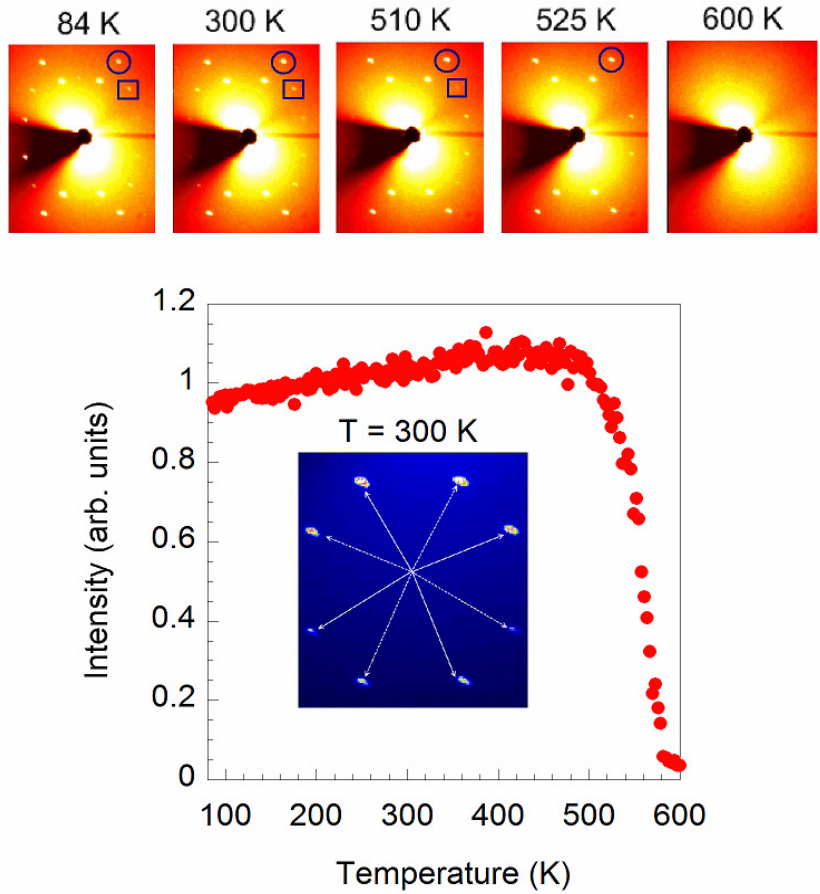

**Figure 2:** (Upper panel) temperature evolution of the spots due to superstructures around the beam-center at five selected temperatures. The first set of superstructure peaks, ($\sqrt{5}\times\sqrt{5}$) are marked with open circles and the second set of weak superstructure peaks ($\sqrt{2}\times\sqrt{2}$) are marked by a squares. Starting high temperature at 600K where there are no superstructure spots, at 525K only the first set of superstructure is detected, and in the frames recorded at 510K and lower temperatures both superstructures are detected. (Lower panel) the intensity of the superstructure peaks (of the first $\sqrt{5}\times\sqrt{5}$ type, marked with circles in the upper panel) during heating run from 80 to 600 K. Inset shows the superstructure peaks (of the first $\sqrt{5}\times\sqrt{5}$ type, marked with circles in the upper panel) as a group of eight spots around the (100) peak position.

Diffraction patterns with superstructure peaks around the beam-center, at five selected temperatures, are shown in the upper panel of figure 2. At 600 K, except for the principal diffraction spots, there are no other superstructure features. The data collected at 525 K show a single group of bright spots in figure 1(a) that can be clearly identified as belonging to the group of eight superstructure reflections around each principal XRD spot (see e.g. the inset in figure 2) identified as the (1/5,3/5,0) diffraction spots in the reciprocal space. The occurrence of well defined eight superstructure spots can be understood as the twin structure described in ref. [18] as described for the TlFe$_{2-x}$Se$_2$ system by Häggström *et al*[20]. The peak intensity variation for one of the superstructure peaks of the first type during the heating run from 80 to 600 K is shown in the lower panel of figure 2. On increasing the temperature, the intensity of these superstructures start to slowly decreases at around 520 K and finally disappears around 580 K.

In addition to the first known set of superstructure peaks, we observe the appearance of a second new set of superstructure peaks below 520 K (see figure 2, upper panel, frame marked as 510 K), assigned to the $\sqrt{2}\times\sqrt{2}$ superstructure.

Figure 3 shows the temperature evolution of the (220) principal diffraction spot. Below 520 K, the peak splits into two components with asymmetric intensities that becomes more evident in the low temperature range. The complete evolution of the (220) peak between 300 to 600 K during the heating run, reconstructed from the CCD image analysis, is shown in the middle panel as a 3D plot of intensity versus temperature and reciprocal lattice wave-vector. A clear peak splitting is evident below 520 K. The relative intensities of the diffraction spots of the two crystal lattices permit to make a rough estimate of the percentage of the two phases. The bottom panel of figure 3 shows the change in the normalized intensity of the majority in-plane expanded phase and the minority in-plane compressed phase. The majority phase has a weight of about 79% at 350 K which becomes 100% above 525 K, with a corresponding vanishing of the intensity of the reflections due to the in-plane compressed minority phase. This data show clearly the intrinsic phase separation in the $K_{0.8}Fe_{1.6}Se_2$ superconductor below 520 K.

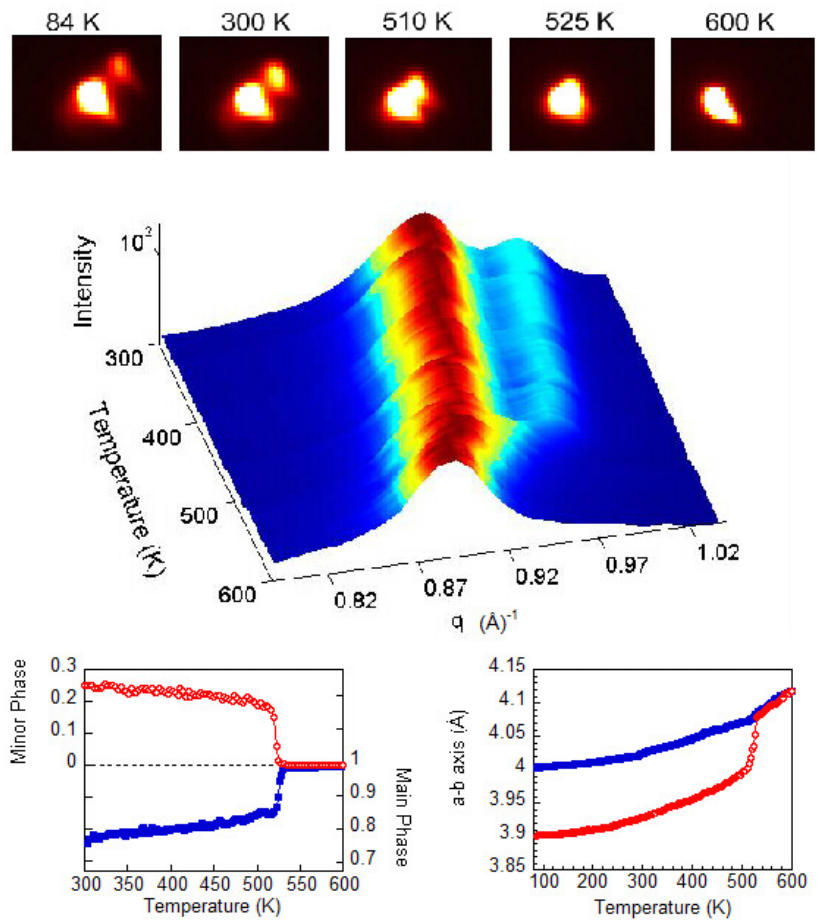

**Figure 3:** (Upper panel) temperature evolution of the (220) peaks at five selected temperatures. (Middle panel) a three dimensional intensity plot of the (220) peak as a function of temperature for the heating run from 300 to 600 K. Below 520 K, the (220) peaks splits, indicating the appearance of a second in-plane *compressed minority phase*. (Lower left-panel) the normalized peak intensity of the first main phase and the second minority phase. (Lower right-panel) the *in-plane* lattice parameter variation with temperature showing the phase separation at 520K.

In Fig. 4, we summarize the results of the present study. During cooling from the disordered single phase at 600 K, the superstructure peaks due to $\sqrt{5}\times\sqrt{5}$ vacancy ordering start to appear at around 580 K, at the same temperature where these peaks disappear in the heating run. In other words, the temperature dependence of the peak intensity variation seems to be identical in the heating and cooling runs.

On the contrary the peak intensity of the (220) satellite during the heating and cooling shows a sharp intensity drop at about 520 K, with a temperature hysteresis of about 10 K. This implies that the intrinsic phase separation occurring at this temperature is primarily of first order nature, unlike the continuous second order nature of the superstructure peaks transition at 580 K. Importantly, the superstructure peak intensity shows no particular change at the phase separation temperature of 520 K. However, as shown in figure 2, a new set of $\sqrt{2}\times\sqrt{2}$ superstructure peaks starts to appear at 520 K. The temperature dependence of the integrated intensities of these new superstructure peaks in a limited temperature range during cooling run is shown in the inset of the left panel in figure 4. The principal diffraction spots of the minority in-plane compressed phase and the new set of $\sqrt{2}\times\sqrt{2}$ superstructure peaks appear at the same temperature, indicating an intimate connection between the two. Further work is going on to characterize this superstructure and it will be object of a longer paper.

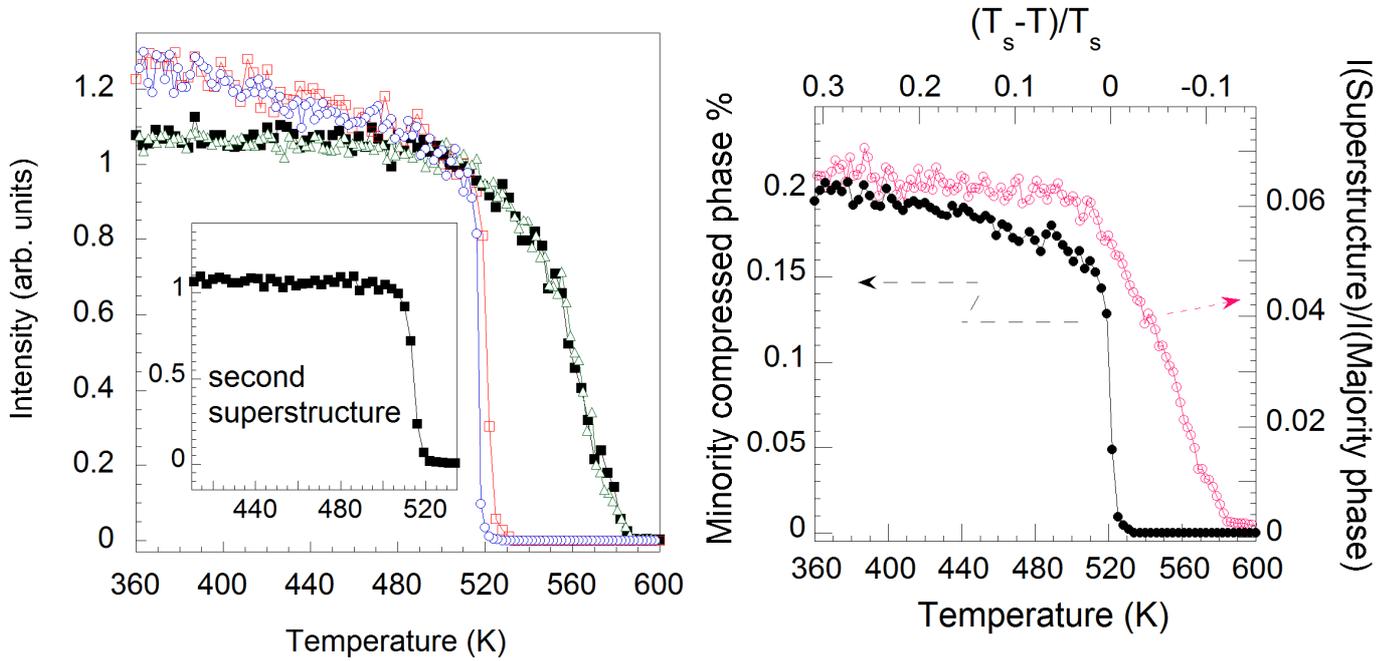

**Figure 4:** (Left panel) the temperature continuous evolution of the $\sqrt{5}\times\sqrt{5}$ superstructure satellite intensity during the heating (open triangles) and cooling (filled squares) runs. The sharp temperature drop of the minority phase during the heating (open squares) and its appearing on cooling (open circles) runs. The inset in the left panel shows the temperature dependence of the integrated intensities of the second weak set of $\sqrt{2}\times\sqrt{2}$ superstructure peaks (filled squares) (refer to the spots in open squares in figure 2). During the cooling run the $\sqrt{2}\times\sqrt{2}$ superstructure peaks show the same temperature onset as the principal XRD diffraction spots of the minority in-plane compressed phase. (Right panel) the variation of the probability of the minority in-plane compressed phase around the phase separation temperature (520 K) (filled dots) and the temperature variation of the ratio (open circles) between the intensity of the $\sqrt{5}\times\sqrt{5}$ iron vacancy superstructure satellites and the principal diffraction spots of the in-plane expanded majority phase.

*Discussion* – We first discuss the ordering of iron vacancies in the majority in-plane expanded phase. The first thing to notice is that the temperature evolution of the associated $\sqrt{5}\times\sqrt{5}$ superstructure spots is in agreement with previous neutron diffraction studies [13,14] on similar systems. The Néel temperature estimated from the neutron diffraction studies is around 560 K, below the order-disorder transition observed at 590 K [13,14], thus, presenting a close analogy between the structural and magnetic transition observed in "1111" family of FeSCs [1,2,21,22] and the new $A_xFe_{2-y}Se_2$ systems. Interestingly, the magnetic and structural transitions are concurrent in the structurally identical "122" pnictides [1,2,23]. As already discussed, the temperature dependence of the appearing and disappearing of the $\sqrt{5}\times\sqrt{5}$ superstructure is continuous without a significant temperature hysteresis indicating the transition to be of second order.

Let us now discuss the new intrinsic phase separation observed here. Earlier structural studies using powder samples [13,14] have not revealed such a transition in $K_{0.8}Fe_{1.6}Se_2$ superconductors. Indeed single-crystal studies provide more insights into the phase separation properties, as revealed in the case of "1111" FeSCs [1,2,21,22]. It is to be stressed that there is a simultaneous sharp appearance of a new set of superstructure peaks together with the minority in-plane compressed phase by decreasing the temperature below 520K.

A recent $^{57}Fe$ Mössbauer study on $K_xFe_{2-y}Se_2$ system revealed abrupt changes in the magnetic fraction of the sample about a similar temperature [24] to the phase separation observed here. An earlier Mössbauer study [20] on $TlFe_{2-y}Se_2$ showed that at lower temperature, the anti-ferromagnetic (AFM) and paramagnetic fractions co-exist with the majority phase to be of AFM nature. The temperature evolution of the magnetic fraction observed in the Mössbauer studies of $K_xFe_{2-y}Se_2$ is also of similar nature. These results indicate the existence of two coexisting phases at low temperature region of the $K_xFe_{2-y}Se_2$. Very recent differential scanning calorimetric measurements on superconducting $Cs_xFe_{2-y}Se_2$ show two distinct heat-flow peaks, larger one at a higher temperature and a weaker one at lower temperature [18], further supporting two possible phases in superconducting $K_xFe_{2-y}Se_2$.

The present observation of the intrinsic phase separation in $K_xFe_{2-y}Se_2$ indicate the importance of lattice complexity of the $A_xFe_{2-y}Se_2$ systems. Indeed the phase diagram involving Fe-Se show the possibility of co-existing phases for different ratio between Fe and Se [25], consistent with extreme sensitivity of the chemical composition to the superconducting properties of FeSe [26]. The coexisting chalcogen heights observed in the doped ternary chalcogenides [27] are found to be more pronounced in the $A_xFe_{2-y}Se_2$ systems [16-18]. The $K_xFe_{2-y}Se_2$ system with its interesting temperature dependent vacancy ordering and competing mesoscopic phases recalls the oxygen ordering effects on the superconductivity of cuprates [28,37]. This intrinsic phase separation is assigned to a generic feature of multi-gaps high temperature superconductors where the system is in the verge of phase separation like in FeAs superlattices [29] and cuprates [30-37]. The data show a clear case of the competition between two electron fluids producing a mesoscopic frustrated phase separation between a majority magnetic metallic phase with $\sqrt{5}\times\sqrt{5}$ ordered defects and a superconducting phase with a second $\sqrt{2}\times\sqrt{2}$ superstructure order with different charge densities; i.e. two striped metallic systems with two different lattice superstructures that will induce different Fermi surface reconstruction with mini-bands and pseudo-gaps of few tens of meV. The present scenario for $A_xFe_{2-y}Se_2$ is similar to the case of overdoped cuprates where two phases compete with different doping [35-38]

***Conclusions*** - We have reported temperature dependent single crystal x-ray diffraction studies of the newly discovered $K_{0.8}Fe_{1.6}Se_2$ superconductor using synchrotron radiation in transmission mode. The basic structure of the sample at room temperature is found to be tetragonal $ThCr_2Si_2$-type, modulated by a vacancy ordering induced superstructures together with a co-existing minority phase with associated different superstructures. The phase separation appears at 520 K, above which the minority in-plane compressed phase merges with the majority in-plane expanded phase. There is a temperature hysteresis of about 10 K, indicating the phase separation transition to be first order. The superstructure peaks corresponding to the main Fe vacancy ordering disappear at 580 K, without noticeable temperature hysteresis, confirming the order-disorder second order phase transition at high temperatures above the phase separation temperature. The present scenario shows that a heterostructure at atomic limit made of a superlattice of metallic layers [39,40] like iron based superconductors tuned at a Lifshitz critical point, [10] is in the verge a lattice catastrophe for phase separation between two types of metals. In the phase separation regime a fine tuning of their Fermiology via two lattice superstructures provides a first itinerant magnetic phase with low lattice symmetry and a second high temperature superconductor tuned at a shape resonance [10] in a landscape of complex phase separation that was called "superstripes" scenario in cuprates [36].

## References

[1] Johnston, D. C. The puzzle of high temperature superconductivity in layered iron pnictides and chalcogenides. *Advances in Physics* **59**, 803-1061 (2010). URL http://dx.doi.org/10.1080/00018732.2010.513480.

[2] Paglione, J. & Greene, R. L. High-temperature superconductivity in iron-based materials. *Nature Physics* **6**, 645-658 (2010). URL http://dx.doi.org/10.1038/nphys1759.

[3] Guo, J. *et al.* Superconductivity in the iron selenide $K_xFe2Se2(0x1.0)$. *Physical Review B* 82, 180520+ (2010). URL http://dx.doi.org/10.1103/PhysRevB.82.180520

[4] Mizuguchi, Y. *et al.* Transport properties of the new fe-based superconductor $K_xFe_2Se_2$ ($T_c$=33K) (2010). URL http://arxiv.org/abs/1012.4950. 1012.4950.

[5] Krzton-Maziopa, A. *et al.* Synthesis and crystal growth of $Cs_{0.8}(FeSe_{0.98})_2$: a new iron-based superconductor with Tc = 27 K. *Journal of Physics: Condensed Matter* **23**, 052203+ (2011). URL http://dx.doi.org/10.1088/0953-8984/23/5/052203.

[6] Shermadini, Z. *et al.* Coexistence of magnetism and superconductivity in the iron-based compound $Cs_{0.8}(FeSe_{0.98})_2$ (2011). URL http://arxiv.org/abs/1101.1873. 1101.1873.

[7] Wang, A. F. *et al.* Superconductivity at 32 K in single-crystalline $Rb_xFe_{2-y}Se_2$. *Physical Review B* **83**, 060512+ (2011). URL http://dx.doi.org/10.1103/PhysRevB.83.060512.

[8] Wang, H.-D. *et al.* 2011 Superconductivity at 32 K and anisotropy in Tl0.58Rb0.42Fe1.72Se2 crystals. *EPL (Europhysics Letters)*47004+ (2011). URL http://dx.doi.org/10.1209/0295-5075/93/47004.

[9] Fang, M. *et al.* Fe-based high temperature superconductivity with $T_c$=31K bordering an insulating antiferromagnet in $(Tl,K)Fe_xSe_2$ crystals (2010). URL http://arxiv.org/abs/1012.5236. 1012.5236.

[10] Innocenti, D. et al. Shape resonance for the anisotropic superconducting gaps near a Lifshitz transition: the effect of electron hopping between layers. Superconductor Science and Technology 24, 015012+ (2011). URL http://dx.doi.org/10.1088/0953-2048/24/1/015012; Innocenti, D. et al. Resonant and crossover phenomena in a multiband superconductor: Tuning the chemical potential near a band edge. Physical Review B 82, 184528+ (2010). URL http://dx.doi.org/10.1103/PhysRevB.82.184528.

[11] Bianconi, A., Agrestini, S., Bianconi, G., Castro, D. & Saini, N. Lattice-Charge stripes in the High-*Tc* superconductors. In *Stripes and Related Phenomena*, Bianconi, A. & Saini, N. L. (eds.) vol. 8 of Selected Topics in Superconductivity, (ISBN 9780306464195) chap. 2, 9-25 (Springer US, Boston, 2000). URL http://dx.doi.org/10.1007/0-306-47100-0_2.

[12] Wang, Z. *et al.* Microstructure and ordering of iron vacancies in the superconductor system $K_yFe_xSe_2$ as seen via transmission electron microscopy. *Physical Review B* **83**, 140505+ (2011). URL http://dx.doi.org/10.1103/PhysRevB.83.140505.

[13] Bao, W. *et al.* A novel large moment antiferromagnetic order in $K_{0.8}Fe_{1.6}Se_2$ superconductor (2011). URL http://arxiv.org/abs/1102.0830. 1102.0830.

[14] Bao, W. *et al.* Vacancy tuned magnetic high-Tc superconductor $K_xFe_{2-x/2}Se_2$ (2011). URL http://arxiv.org/abs/1102.3674. 1102.3674.

[15] Zavalij, P. *et al.* On the structure of vacancy ordered superconducting potassium iron selenide (2011). URL http://arxiv.org/abs/1101.4882. 1101.4882.

[16] Bacsa, J. *et al.* 2011 Cation vacancy order in the K0.8+xFe1.6-ySe2 system: Five-fold cell expansion

accommodates 20% tetrahedral vacancies. *Chem. Sci.* 2, 1054. URL http://dx.doi.org/10.1039/c1sc00070e.

[17] Wang, D. M., He, J. B., Xia, T. L. & Chen, G. F. Effect of varying iron content on the transport properties of the potassium-intercalated iron selenide $k_x$fe$_2-y$se$_2$. *Physical Review B* **83**, 132502+ (2011). URLhttp://dx.doi.org/10.1103/PhysRevB.83.132502.

[18] Yu V *et al.* Iron-vacancy superstructure and possible room-temperature antiferromagnetic order in superconducting cs$_y$fe$_2-x$se$_2$. *Physical Review B* **83**, 144410+ (2011). URL http://dx.doi.org/10.1103/PhysRevB.83.144410.

[19] Ye, F. *et al.* Common structural and magnetic framework in the $A_2Fe_4Se_5$ superconductors (2011). URLhttp://arxiv.org/abs/1102.2882. 1102.2882.

[20] Häggström, L. A Mössbauer study of antiferromagnetic ordering in iron deficient $TlFe_{2-x}Se_2$. *Journal of Magnetism and Magnetic Materials* **98**, 37-46 (1991). URL http://dx.doi.org/10.1016/0304-8853(91)90423-8.

[21] Ricci, A., Fratini, M. & Bianconi, A. The tetragonal to orthorhombic structural phase transition in multiband FeAs-based superconductors. *Journal of Superconductivity and Novel Magnetism* **22**, 305-308 (2009). URLhttp://dx.doi.org/10.1007/s10948-008-0434-9.

[22] Ricci, A. *et al.* Structural phase transition and superlattice misfit strain of RFeAsO (R=La, Pr, Nd, Sm). *Physical Review B* **82**, 144507+ (2010). URL http://dx.doi.org/10.1103/PhysRevB.82.144507. ; Jesche, A., Krellner, C., de Souza, M., Lang, M. & Geibel, C. Coupling between the structural and magnetic transition in CeFeAsO. *Physical Review B* **81**, 134525+ (2010). URL http://dx.doi.org/10.1103/PhysRevB.81.134525. ; Tian, W. *et al.* Interplay of Fe and Nd magnetism in NdFeAsO single crystals. *Physical Review B* **82**, 060514+ (2010). URLhttp://dx.doi.org/10.1103/PhysRevB.82.060514.

[23] Rotter, M. *et al.* Spin-density-wave anomaly at 140 k in the ternary iron arsenide $BaFe_2As_2$. *Physical Review B* **78**, 020503+ (2008). URL http://dx.doi.org/10.1103/PhysRevB.78.020503.

[24] Ryan, D. H. *et al.* $^{57}$Fe mössbauer study of magnetic ordering in superconducting $K_{0.80}Fe_{1.76}Se_2$ single crystals. *Physical Review B* **83**, 104526+ (2011). URL http://dx.doi.org/10.1103/PhysRevB.83.104526.

[25] Predel, B.: 1995 *Fe-Se (Iron-Selenium)*. Madelung, O. (ed.). SpringerMaterials - The Landolt-Börnstein Database DOI: 10.1007/10474837_1339 URL http://www.springermaterials.com/index/bookdoi:10.1007/b55397

[26] McQueen, T. M. *et al.* Extreme sensitivity of superconductivity to stoichiometry in $Fe_{1+x}Se$. *Physical Review B* **79**, 014522+ (2009). URL http://dx.doi.org/10.1103/PhysRevB.79.014522.

[27] Joseph, B. *et al.* Evidence of local structural inhomogeneity in $FeSe_{1-x}Te_x$ from extended x-ray absorption fine structure. *Physical Review B* **82**, 020502+ (2010). URL http://dx.doi.org/10.1103/PhysRevB.82.020502.

[28] Fratini, M. *et al.* Scale-free structural organization of oxygen interstitials in $La_2CuO_{4+y}$. *Nature* **466**, 841-844 (2010). URLhttp://dx.doi.org/10.1038/nature09260

[29] Caivano, R. et al. Feshbach resonance and mesoscopic phase separation near a quantum critical point in multiband FeAs-based superconductors. Superconductor Science and Technology 22, 014004+ (2009). URL http://dx.doi.org/10.1088/0953-2048/22/1/014004.

[30] Gorkov, L. & Teitelbaum, G. Pseudogap regime in high- cuprates as a manifestation of a frustrated phase separation (NMR view). Physica B: Condensed Matter 359-361, 509-511 (2005). URL http://dx.doi.org/10.1016/j.physb.2005.01.130.

[31] Dagotto, E. Complexity in Strongly Correlated Electronic Systems. *Science* **309**, 257-262 (2005).

[32] de Mello, E. V. L. & Dias, D. H. N. Phase separation and the phase diagram of cuprate superconductors. Journal of Physics: Condensed Matter 19, 086218+ (2007). URL http://dx.doi.org/10.1088/0953-8984/19/8/086218.

[33] Bishop, A. R. HTC oxides: a collusion of spin, charge and lattice. J. Phys. Conf. Ser.108, 012027 (2008). doi: 10.1088/1742-6596/108/1/012027

[34] Müller K. A. On the superconductivity in hole doped cuprates *Journal of Physics: Condensed Matter* 19, 251002 (2007). doi: 10.1088/0953-8984/19/25/251002

[35] Innocenti, D. et al. A model for Liquid-Striped liquid phase separation in liquids of anisotropic polarons. Journal of Superconductivity and Novel Magnetism 22, 529 (2009). URL http://dx.doi.org/10.1007/s10948-009-0474-9.

[36] Bianconi, A. Superstripes. International Journal of Modern Physics B 14, 3289-3297 (2000). URL http://dx.doi.org/10.1142/S0217979200003769.

[37] Poccia, N., Ricci, A. & Bianconi, A. Fractal structure favoring superconductivity at high temperatures in a stack of membranes near a strain quantum critical point. Journal of Superconductivity and Novel Magnetism 24, 1195-1200 (2011). URL http://dx.doi.org/10.1007/s10948-010-1109-x.

[38] Kugel, K. I., Rakhmanov, A. L., Sboychakov, A. O., Poccia, N. & Bianconi, A. Model for phase separation

controlled by doping and the internal chemical pressure in different cuprate superconductors. *Physical Review B* **78**, 165124+ (2008). URL http://dx.doi.org/10.1103/PhysRevB.78.165124.

[39] Bianconi, A. On the possibility of new high Tc superconductors by producing metal heterostructures as in the cuprate perovskites. *Solid State Communications* **89**, 933-936 (1994). URL http://dx.doi.org/10.1016/0038-1098(94)90354-9.

[40] Ricci, A. *et al.* On the possibility of a new multiband heterostructure at the atomic limit made of alternate $CuO_2$ and FeAs superconducting layers. *Superconductor Science and Technology* **23**, 052003+ (2010). URL http://dx.doi.org/10.1088/0953-2048/23/5/052003.